\documentclass{article}

\usepackage{arxiv}

\usepackage[utf8]{inputenc} 
\usepackage[T1]{fontenc}    
\usepackage{hyperref}       
\usepackage{url}            
\usepackage{booktabs}       
\usepackage{amsfonts}       
\usepackage{nicefrac}       
\usepackage{microtype}      
\usepackage{lipsum}		
\usepackage{graphicx}
\usepackage{natbib}
\usepackage{doi}
\usepackage{amsmath}
\usepackage{authblk}

\title{A Unified Dynamic Model for the Decomposition of Skin Conductance and the Inference of Sudomotor Nerve Activities
\thanks{\textit{\underline{Citation}}: 
\textbf{This work has been submitted to the IEEE for possible publication. Copyright may be transferred without notice, after which this version may no longer be accessible.}} 
}

\author[1]{Hui Sophie Wang}
\author[1, 2]{Stacy Marsella}
\author[1, 3]{Misha Pavel}

\affil[1]{Khoury College of Computer Science, Northeastern University, Boston, MA 02115, USA}
\affil[2]{Department of Psychology, Northeastern University, Boston, MA 02115, USA}
\affil[2]{Bouvé College of Health Sciences, Northeastern University, Boston, MA 02115, USA}
\affil[ ]{\{wang.hui1, s.marsella, m.pavel\}@northeastern.edu}

\date{}

\begin{document}
\maketitle

\begin{abstract}
In behavioral health informatics, inferring an individual's psychological state from physiological and behavioral data is fundamental. A key physiological signal in this endeavor is electrodermal activity (EDA), often quantified as skin conductance (SC), known for its sensitivity to a variety of psychological stimuli. Traditional methods to analyze skin conductance, such as the trough-to-peak method, often result in imprecise estimations due to overlapping skin conductance responses. While various mathematical models have been proposed to improve the analysis, many of them do not incorporate the tonic level in the dynamic system. This paper introduces a novel fourth order dynamic system to model the temporal dynamics of skin conductance, unifying both the tonic level and phasic response. Applied to a large dataset with over 200 participants, majority of the models achieved an $R^2$ value above 0.99. Furthermore, this work offers a unique three-component decomposition of skin conductance, shedding light on its temporal dynamics. Comparative evaluations highlight the model's capability to differentiate arousal levels and maintain an appropriate sparsity level for the estimated sudomotor nerve activities signal. The code of the proposed model and algorithm are available for open access. 
\end{abstract}


\section{Introduction}
\label{sec:introduction}

In the field of behavioral health informatics, inferring an individual's psychological states from physiological and behavioral data is an important problem that lies at the core of many applications. One of the most widely used physiological signals is electrodermal activity (EDA), given its sensitivity to an extensive range of psychological stimuli \cite{dawson2017}. EDA refers to the electrical properties of the skin \cite{boucsein2012}. One of the most important factors that influence the electrical properties of the skin is sweat secretion. Sweat glands are innervated by the sympathetic nervous system, a branch of the autonomic nervous system that controls the body’s “fight-or-flight” response \cite{cannon1939}. Therefore, EDA is closely linked with physiological and psychological arousal.

There exist several methods to measure EDA, the most prevalent being skin conductance (SC), which is assessed by passing a small current between a pair of skin surface electrodes \cite{dawson2017}. Skin conductance is assumed to comprise two components—tonic level (SCL) and phasic response (SCR). The tonic level is defined as the background level in the absence of any internal or external stimuli, while the phasic response denotes responses to stimuli \cite{stern2001}. Historically, the trough-to-peak method has been used to analyze EDA \cite{benedek2010_CDA}. This approach detects local troughs and peaks using a minimum threshold, identifying an SCR by a pair of trough and peak. However, because SCRs often overlap with each other, this method tends to yield inaccurate estimations of SCR amplitudes and frequency \cite{boucsein2012}.

Compared to the traditional approach, mathematical and computational models hold distinct advantages in inferring latent psychological variables from observable physiological measurements. The relationship between psychological variables and physiological measurements is intricate and context-dependent. Non-modeling approaches, or more accuratly implicit models, lack clearly specified assumptions, leading to significant ambiguity and making evaluation challenging \cite{epstein2008, bach2013model}.  In contrast, mathematical models offer precise definitions of the constructs and clearly lay out model assumptions, allowing for thorough examination and evaluation of the model. These models also hold the potential to provide insights into the phenomena by striving to characterize the fundamental processes and underlying mechanisms. Moreover, mathematical models can be used for simulation, allowing for the identification of limitations and the derivation of new hypotheses for future explorations.

Several mathematical approaches have been proposed to model EDA, including \cite{lim1997, alexander2005, benedek2010_CDA,benedek2010_DDA, bach2010modelling, bach2010dynamic, greco2015, hernando2017, amin2020}. One issue of these existing models is that the tonic level is not defined rigrously and thus not incorporated into the dynamic models. In most of the models SCL is estimated by cubic splines \cite{benedek2010_CDA,benedek2010_DDA, greco2015, hernando2017, amin2020}. This method necessitates setting a value for the spacing of the cubic splines' fixed points. However, finding an appropriate value can be challenging, as it may vary based on the frequency of SCRs. Consequently, constant spacing sometimes leads to arbitrary dynamics outside the fixed points. In addition, any residual errors of the model can be accommodated by the cubic splines because of their flexibility. Another method involves removing SCL using a high-pass filter \cite{bach2010modelling, bach2010dynamic}, but this is problematic due to the likely overlap of frequency bands of tonic and phasic components, leading to signal distortion and information loss about the shape of the SCRs.

To address this issue, this work proposes a novel unified dynamic model that combines the tonic level and phasic response dynamics within a single dynamic system.  Amin and Faghih recently proposed BayesianEDA \cite{amin2022}, which also integrates these two components within a single dynamic system. Developed independently, the model proposed here is very different from BayesianEDA in terms of key model assumptions. The shared interest in this problem reflects the importance of this issue. Addressing this issue is essential for improving EDA analysis and deepening our understanding of the relationships between psychological states and EDA metrics. The shared interest in this problem underscores its significance. Addressing it is essential for improving EDA analysis and elucidating the relationships between psychological states and EDA metrics.

The remainder of this paper is organized as follows: Section II reviews the assumptions behind existing EDA models. In Section III, we describe the proposed model in detail, highlighting its structure, variables, and key assumptions. Section IV presents the algorithm for estimating model parameters and latent variables. The results of model selection and fitting are summarized in Section V. Section VI describes skin conductance decomposition. Section VII focuses on the assessment of the estimated sudomotor nerve activities, a neural signal response for skin conductance responses. We conclude with Section VIII, and Section IX follows with discussions.

\section{Related Works}
Several mathematical approaches have been proposed to model electrodermal responses, including \cite{lim1997, alexander2005, benedek2010_CDA,benedek2010_DDA, bach2010modelling, bach2010dynamic, greco2015, hernando2017, amin2020, amin2022}. These approaches can be compared on several aspects. One of the key aspects is the model assumptions. Many of the models share similar assumptions that are based on physiology. The secretory part of sweat glands receives direct innervation from postganglionic sudomotor nerve fibers, a type of sympathetic nerve fibers. These sudomotor nerve activities (SNA) are recognized as the cause of the phasic aspect of EDA, as supported by microneurography studies \cite{bini1980, wallin1981}. Based on this causal relationship, the phasic response is often modeled by a Linear time-invariant (LTI) dynamic system with SNA as input and SCR as output. The SNA signal is often assumed to be sparse or compact, based on the bursting pattern exhibited by large numbers of neighboring sudomotor neurons firing synchronously \cite{mcallen1997}. The mechanism of the tonic level is less well understood. It was suggested that tonic level can be attributable to corneal hydration, which can be impacted by many factors such as ambient temperature and skin temperature \cite{boucsein2012}. Therefore, most models did not make explicit assumptions about the tonic level. However, the mathematical choices and realization may reflect implicit assumptions. For example, the spacing of the fixed points of cubic splines used for interpolating the tonic level indicates implicit assumptions on how fast the tonic level changes. Table \ref{model_assumptions} contains the assumptions of the existing models, including explicit assumptions as well as the mathematical realizations related to implicit assumptions of the models, regarding the tonic level, phasic response and sudomotor nerve activities.

\begin{table*}[t]
\caption{Assumptions of Existing EDA Models}

\begin{center}
\begin{tabular}{ |p{20mm}|p{45mm}|p{45mm}|p{45mm}| }
\hline
\textbf{Model} & \textbf{Tonic Level} & \textbf{Phasic Response} & \textbf{Sudomotor Nerve Activities} \\
\hline
Lim et al. \cite{lim1997} & 
Constant for each 10 s segment & 
Sigmoid-exponential function was used to fit each SCR. & NA \\
\hline
Alexander et al. \cite{alexander2005} & NA & 
LTI system with bi-exponential function as impulse response. & 
Sudomotor bursts are discrete, separated events, whose time constant is much smaller than SCR.\\
\hline
Ledalab CDA \cite{benedek2010_CDA} & Interpolated by cubic splines with fixed points spaced every 10 s & LTI system with bi-exponential function as impulse response. & Distinct and compact impulses with short support (nonzero values) \\
\hline
Ledalab DDA \cite{benedek2010_DDA} & Interpolated by cubic splines with fixed points spaced every 100 s & LTI system with bi-exponential function as impulse response. & Nonnegative with maximal compactness \\
\hline
PsPM GLM \cite{bach2010modelling} & Removed by a high pass filter with cutoff frequency at 0.05 Hz & LTI system with Gaussian smoothed bi-exponential function as canonical response function. Each SCR is a linear combination of the canonical response function and its derivatives. & Not modeled explicitly. The amplitude of each SNA burst is assumed to be the amplitude of the corresponding SCR. \\
\hline
PsPM DCM \cite{bach2010dynamic} & Removed by a high pass filter with cutoff frequency at 0.0159 Hz (corresponding to a time constant of 10 s) & Nonlinear state space model with canonical response functions derived from previous data & Each SNA burst is represented by a Gaussian function with varying amplitudes but fixed timing and duration. \\
\hline
cvxEDA \cite{greco2015} & Linear trends and cubic B-splines with fixed points spaced every 10 s & LTI system with bi-exponential function as impulse response. & Nonnegative and sparse. \\
\hline
sparsEDA \cite{hernando2017} & An unspecified function and its first order Taylor series expansion & LTI system with multi-scale bi-exponential function as response functions. & Nonnegative and as sparse as possible. \\
\hline
Amin and Faghih \cite{amin2020} & Interpolated by cubic B-splines with fixed points spaced every 6 s & LTI system with bi-exponential function as impulse response. & Nonnegative and sparse. \\
\hline
Bayesian EDA \cite{amin2022} & Tonic level is caused by the sweat in the stratum corneum, represented as a state in a 3D state space model. & Phasic response is caused by the sweat pushed out to the surface of the skin through pore opening, represented as a state in a 3D state space model. & Sparse. The amplitudes of SNA bursts are assumed to follow generalized Gaussian distribution, and are above a minimum threshold of 0.25 $\mu S/s$. \\
\hline
\end{tabular}
\end{center}
\label{model_assumptions}
\end{table*}

\section{Model Description}

\subsection{Overall Model Structure}

The overall structure and components of the proposed model are represented by the block diagram in the figure below. The input of the system is stimuli $u(t)$, which can be external events or internal thoughts and feelings. In the broader context, $u(t)$ is typically a multi-dimensional signal. Without loss of generality, in this work, we consider it as one-dimensional. The output of the system is $SC(t)$, which is a one-dimensional signal that represents the skin conductance measurement. 

A key variable in this model is the sudomotor nerve activities (SNA), represented as a one-dimensional signal $v(t)$. Although the time variable is assumed to be a continuous function of time, we represent it by discrete samples with a sampling interva 0.1 sec. SNA is the neural signal that generates SCR, the phasic component of $SC(t)$. In this work, $v(t)$ is treated as a latent variable, since there is no direct measurement of it. We represent $v(t)$ using a series of impulse clusters, with each cluster representing a burst. A impulse cluster consists of several adjacent impulses with non-negative amplitudes. A single impulse is a time-shifted and scaled dirac delta function in discrete time. The choice of representing each SNA burst as a cluster of impulses instead of a single impulse was informed by microneurography studies, which revealed that the mean duration of sudomotor bursts is around 0.6 seconds \cite{bini1980, macefield1996}. Given the $v(t)$ signal's sampling interval of 0.1 seconds (at 10 Hz), employing a cluster of adjacent impulses appears more realistic and better equipped to capture the diverse shapes and durations observed in the SNA signal. Given a sampling interval of 0.1 seconds for $v(t)$, a cluster of adjacent impulses provides a more realistic representation and allows us to effectively estimate the varying shapes and durations of the SNA bursts.

The main components of the model are the neural system and the peripheral system. The neural system represents the overall process/function of the nervous system that transforms the stimuli $u(t)$ into the sudomotor nerve activities signal $v(t)$. The peripheral system represents the transformation from $v(t)$ to $SC(t)$ in the periphery. It should be noted that although this diagram is based on biological processes, it is a much simplified representation to provide the appropriate level of abstraction for this modeling work. For example, no feedback loops are included. In addition, the neural system encapsulates the complex physiological and psychological processes underlying the perception of the stimuli. The physiological and psychological processes are not viewed separately, but as one integrated complex process.

\begin{figure}[th]
\centering
\vspace{0.3cm}
\includegraphics[scale=0.5]{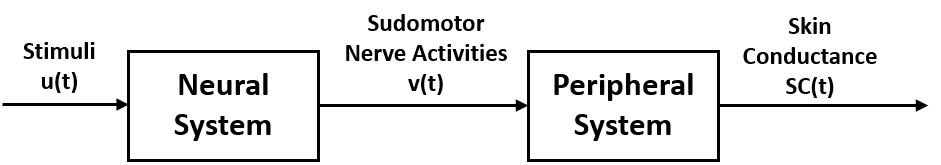}
\vspace{0.5cm}
\caption{Overall Model Structure}
\label{fig:overall}
\end{figure}

\subsection{Model Assumptions of the Peripheral System}

The focus of this work is the peripheral system. The proposed model of the peripheral system is referred to as the unifed dynamic model (UDM). The following list of assumptions define the UDM.

The first assumption is that the dynamics of the peripheral system can be described by an nth order discrete linear time-invariant system with an additive noise term $e(t)$. The model's prediction or approximation of skin conductance is denoted as $y(t)$, which represents the measurement $SC(t)$ with noise $e(t)$ subtracted from it. Noise $e(t)$ comprises the measurement noise and the part of the peripheral system that is not captured by the LTI system. The LTI assumption permits us to represent the relationship between $v(t)$ and $y(t)$ by Equation \ref{eq:difference_eq1}. One interpretation of this equation is that $y(t)$, the output of the system at time $t$, is a linear combination of the past states  and inputs.

\begin{equation}
y(t) +a_1y(t-1)+...+a_ny(t-n) = b_0v(t)+b_1v(t-1)+...+b_nv(t-n)
\label{eq:difference_eq1}
\end{equation}

The order $n$ of the LTI system is not assumed at this point, but it is expected that $n$ is equal to or larger than two. Many of the existing EDA models use the bi-exponential function as the impulse response function for the phasic response, which is equivalent to a second order LTI system. Since the proposed model aims to capture the dynamics of both the tonic level and the phasic response, an order of at least two is expected.

The second assumption is about the decomposition of the LTI system's output $y(t)$ into the tonic level and the phasic response. Our assumption is that the tonic level is independent of the inputs and can be represented by the free response $y_{free}(t)$. The phasic component can be represented by forced response $y_{forced}(t)$ of the LIT system. The free response refers to the output of the system under nonzero initial condition and zero input. The initial condition is the initial state of the dynamic system at the time $t=0$. For a dynamic system of order $n$, the initial state is an n-dimensional vector. Therefore, $y_{free}(t)$ can be obtained by setting the input $v(t)$ to zero in Equation \ref{eq:difference_eq1} and advancing time $t$. On the other hand, the forced response represents the output of the system under zero initial condition and nonzero input, and $y_{forced}(t)$ can be obtained by setting the initial state of the system to zero and advancing time $t$. The output of the system $y(t)$ is the sum of the free response and the forced response, as described by Equation \ref{eq:difference_sum}.

\begin{equation}
y(t) = y_{free}(t) + y_{forced}(t)
\label{eq:difference_sum}
\end{equation}

This assumption combines the dynamics of the tonic level and the phasic response into one dynamic system. Through this assumption, the output of the dynamic system is decomposed into two independent components, one component (phasic response) is attributable to the external input, and the other (tonic level) is attributable to the initial state of the system. Thus the phasic response is defined as the electrodermal response to any known or unknown stimuli. Nonzero initial state represents the dynamics of SCR that are independent of the immediate inputs. Skin conductance rarely reaches zero even in the absence of external input. The nonzero initial state can be viewed as summarizing the state of the system determined by a history of the neural activities and other factors up until the present time.

The third assumption is that sudomotor nerve activities v(t), the neural signal that generates the phasic response, is a sparse and nonnegative signal.  A sparse signal has a majority of zero values with some nonzero values dispersed among the zero values. The nonzero values represent bursts, which correspond to high level neural activities, and the zero values represent low level neural activities. The sparsity assumption or similar assumptions of the neural signal was made by almost all of the existing EDA models. The sparsity assumption not only captures the physiological property of the SNA signal, it also allows us to frame the problem of estimating $v(t)$ as a sparse representation problem, facilitating the use of various compressed sensing techniques \cite{zhang2015}.

The fourth assumption is that the parameters of the dynamic model vary among individuals. This assumption is based on the fact that electrodermal responses exhibit considerable individual differences. For instance, electrodermal labiles refer to individuals who show high rates of nonspecific skin conductance responses at resting conditions and/or slow habituation to repeated stimuli \cite{lacey1958}. Electrodermal lability is shown to be a stable trait with high test-retest reliability, and is found to be associated with behavioral differences and personality characteristics \cite{dawson2017}. In addition, demographic factors such as age and gender also influence EDA responses \cite{boucsein2012}. This assumption allows the model to capture such individual differences.  

To summarize, these are the key assumptions made by the proposed model:

\textbf{A1.} The transformation from sudomotor nerve activity $v(t)$ to skin conductance $SC(t)$ can be represented by an linear time invariant dynamic system. 

\textbf{A2.} The tonic level and phasic response can be represented respectively by the free response $y_{free}(t)$ and the forced response $y_{forced}(t)$ of the dynamic system. Both the tonic level and phasic response should be nonnegative.

\textbf{A3.} Sudomotor nerve activities $v(t)$ is a sparse and nonnegative signal. 

\textbf{A4.} The parameters of the dynamic system vary among individuals.

\section{Model Estimation Algorithm}
\label{sec_model_est}

In this section, we describe the algorithm used to estimate model parameters and the latent variable $v(t)$. The need to estimate the system parameters and $v(t)$ required us to develop an interative algorithm which is a variation of the Expectation Maximization (EM) algorithm \cite{moon1996}. EM is an iterative method that obtains locally optimal model parameters in the presence of latent variables. EM typically starts with an initial estimation of the model parameters, and alternates between an Expectation step (E-step) and a Maximization step (M-step). The code of the model and the algorithm are available at \url{https://github.com/huisophiewang/UDM_EDA}

\subsection{Initialization}

In this work, EM algorithm starts with an initial estimation of the latent variables instead of the model parameters. The reason of this choice is due to a lack of prior knowledge about the ranges and distributions of the model parameters of the proposed model. This model differs from existing models in several key aspects, including nonzero initial conditions and higher model order, which makes the ranges and distributions of the model parameters of existing models not applicable. On the other hand, an initial estimation of the latent variables v(t) can be obtained using the trough-to-peak method traditionally used to obtain SCR amplitudes \cite{benedek2010_CDA}. In the trough-to-peak method, pairs of local valleys and peaks obtained through peak detection approximate the onsets and peaks of SCRs. The amplitude of an SCR is estimated by the height difference of the onset and peak. The initial $v(t)$ consists of a series of single impulses located at the valleys. A threshold of 0.01 microSiemens was used as the minimum amplitude \cite{boucsein2012}.

\subsection{Estimating Model Parameters}
\label{sec_M_step}

This section describes the M-step of the algorithm, in which the model parameters are estimated assuming the latent variable $v(t)$ known. In addition to the model parameters, the initial condition of the model is also unknown and needs to be estimated. We used Matlab’s System Identification Toolbox \cite{ljung1995} in this step, which supports the estimation of the initial condition together with model parameters. Different model structures were explored, and transfer function model was chosen based on better fitting results than ARX model and state-space model. The implementation of these models in Matlab differ in the noise term and the optimization method used, although they are essentially interchangeable models. 

An important hyper-parameter is the model order. As model order increases, the model’s complexity increases, so is the risk of overfitting. Selecting an appropriate model order is essential to prevent underfitting or overfitting. Because the process of model selection involves analyzing data, it will be discussed in detail in Section \ref{sec_model_sel}. For the purpose here, it can be assumed that model order is selected.

An LTI system can be decomposed into the sum of simpler subsystems with one real pole or a pair of complex poles through partial fraction decomposition of its transfer function. The subsystems have simpler forms and their behaviors are much easier to understand intuitively. A better understanding of the whole system can be obtained from the subsystems. We implemented a pruning process based on this decomposition. Because skin conductance is a relatively slow-varying signal, complex-pole subsystems with a period larger than one second are likely to be fitting noise. By eliminating these subsystems and summing up the remaining ones, we obtain a reconstructed model that is more parsimonious and robust.

\subsection{Estimating Latent Variables}
\label{sec_E_step}

This section describes the E-step of the algorithm, in which the latent variables are estimated assuming the model parameters are known. Suppose both $y(t)$ and $v(t)$ have N data points at $t=0,1,…,N-1$, a list of equations of the same form as Equation \ref{eq:difference_eq1} can be written for each data point at $t=n-1,n,…,N-1$ ($n$ is the model order). Stacking these $N-n+1$ equations vertically forms the following equation in matrix form:

\begin{equation}
\textbf{A} \textbf{y} = \textbf{B} \textbf{v} \\
\label{eq:Ay_Bv}
\end{equation}

where \textbf{A} and \textbf{B} are matrices whose elements are the parameters of the model. The dimension of \textbf{A} and \textbf{B} are $(N-n) \times N$. Both \textbf{y} and \textbf{v} are column vectors of length N. 

Because $y(t)$ can be approximated by $SC(t)$, $v(t)$ is the only unknown variable in Equation \ref{eq:Ay_Bv} and can be obtained by solving the system of linear equations. Since there are fewer equations ($N-n+1$) than the number of variables ($N$), the system is under-determined, and therefore has an infinite number of solutions. Among all the possible solutions, the solution that minimizes the distance between $y(t)$ and $SC(t)$ is preferred. Therefore, solving for $v(t)$ can be formulated as an optimization problem:

\begin{equation}
\begin{aligned}
\min_{v} \quad & {\Vert y(t) - SC(t) \Vert}^2_2 \\
\label{eq:opt1}
\end{aligned}
\end{equation}

Substituting $\textbf{A}^{-1} \textbf{B} \textbf{v}$ for $y(t)$ and representing SC(t) with vector $\textbf{c}$, the optimization \ref{eq:opt1} becomes:

\begin{equation}
\begin{aligned}
\min_{v} \quad & {\Vert {\textbf{A}^{-1} \textbf{B} \textbf{v} - \textbf{c}} \Vert}^2_2 \\
\label{eq:opt2}
\end{aligned}
\end{equation}

Note that \textbf{A} is not a square matrix and therefore is not invertible. $\textbf{A}^{-1}$ can be obtained through pseudoinverse or matrix left division. 

In addition, $v(t)$ needs to be sparse and nonnegative according to model assumption A3. The nonnegative assumption can be easily incorporated as a constraint. The sparsity assumption can be formulated as $L_1$ norm regularization, which is one of the most widely used sparse representation techniques. The optimization becomes:

\begin{equation}
\begin{aligned}
\min_{v} \quad & {\Vert {\textbf{A}^{-1} \textbf{B} \textbf{v} - \textbf{c}} \Vert}^2_2 + \alpha {\Vert \textbf{v} \Vert}_1 \\
\textrm{s.t.} \quad & \textbf{v} \geq0
\label{eq:opt4}
\end{aligned}
\end{equation}

Finally, this optimization problem is equivalent to the following one, which can be solved efficiently by standard quadratic programming:
 
\begin{equation}
\begin{aligned}
\min_{v} \quad & \frac{1}{2} \textbf{v}^{T} \textbf{D}^{T} \textbf{D} \textbf{v} + (-\textbf{D}^{T}\textbf{c} + \alpha) \textbf{v}  \\
\textrm{s.t.} \quad & \textbf{v} \geq0
\label{eq:opt6}
\end{aligned}
\end{equation}

where $\textbf{D} = \textbf{A}^{-1}\textbf{B}$ is a square matrix of dimension $N \times N$.

\section{Model Selection and Fitting}
\label{sec_model_sel}

Model selection in this context refers to the process of determining the order $n$ of Equation \ref{eq:difference_eq1}. In order to avoid overfitting, it is common in machine learning to choose the best hyper-parameters using validation data. However, this approach is not suitable here, because we are estimating a model for each individual, and the data points in a skin conductance recording cannot be considered as independent and identically distributed. In fact, one of the purposes of this work is to investigate the temporal dynamics of skin conductance. Therefore, other techniques are used instead in this work to reduce the chance of overfitting, including pruning and regularization (described in Section \ref{sec_model_est}).

The dataset used in this section (referred to as IASL\_ARB) contains 260 participants who completed an acoustic startle task. During the task, 24 trials of loud white noise (80 db with instantaneous rise time) were played over headphones in randomly controlled intervals varied from 10-20 seconds. Participants were not required to perform anything other than listening through the headphone. The total duration of the startle task was around 10 minutes, including a 3-minute rest period before the first trial starts. This dataset was collected as part of a larger study in which participants completed a number of other tasks unrelated to the current investigation. We chose this dataset because it contains a relatively large number of participants and its stimuli has well-defined temporal definition.

\subsection{Preprocessing}
\label{sec_prep}

For data pre-processing, a low-pass filter was applied to the raw skin conductance signal SC(t) to reduce the higy-frequency noise. A second order Butterworth filter was used with a cutoff frequency of 5 Hz. Because skin conductance is relatively slow varying, this cutoff frequency is high enough to preserve the signal. Next, the filtered SC(t) was downsampled to 10 Hz, and trimmed to contain the segment from the end of the resting period to the end of the last trial.

Thirty-six participants were excluded from analysis due to invalid data, including 15 participants with missing skin conductance recording files or timing information, 15 participants with whole sessions or segments of unchanging skin conductance at the minimum level, 6 participants whose recordings have sudden large drops that might be caused by electrodes temporarily losing contact with the skin. In addition, 4 participants were excluded because their skin recordings only contain the tonic level without any detected skin conductance responses (the minimum amplitude threshold of 0.01 $\mu S$ was used). In sum, 220 participants ($86.2\%$) were used for model estimation.

\subsection{Procedure and Results}

\begin{figure}[!ht]
\centering
\includegraphics[scale=0.5]{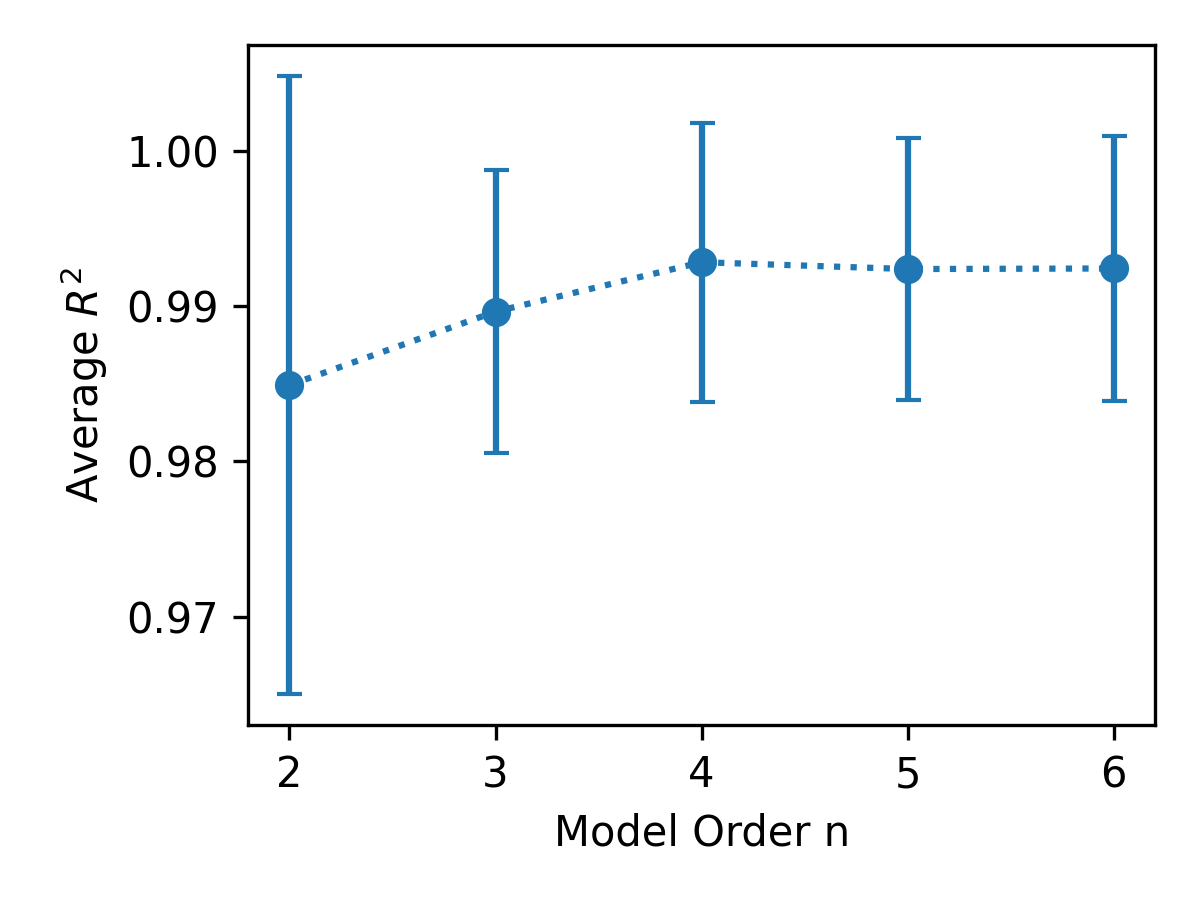}
\caption{Selecting Model Order Based on Average $R^2$}
\label{fig:model_order}
\end{figure}

The following procedure was used for selecting the best model order $n$. The model order $n$ was set to vary from 2 to 6. For each $n$, a model of order $n$ is estimated from each individual’s skin conductance using the model estimation algorithm described in Section \ref{sec_model_est}. The goodness of fit of the model is evaluated by $R^2$, which is defined as the percentage of variance explained by the model. $R^2$ is then averaged across each individual’s estimated model. Figure \ref{fig:model_order} shows the model order against the average $R^2$. The average $R^2$ is highest when the model order is four. Increasing the model order beyond four doesn't improve the average $R^2$. For fourth-order models, $95.5\%$ (210 out of 220) have an $R^2$ value above 0.99, suggesting that the proposed model fits the observed data well.

\begin{figure}[!ht]
\centering
\includegraphics[scale=0.23]{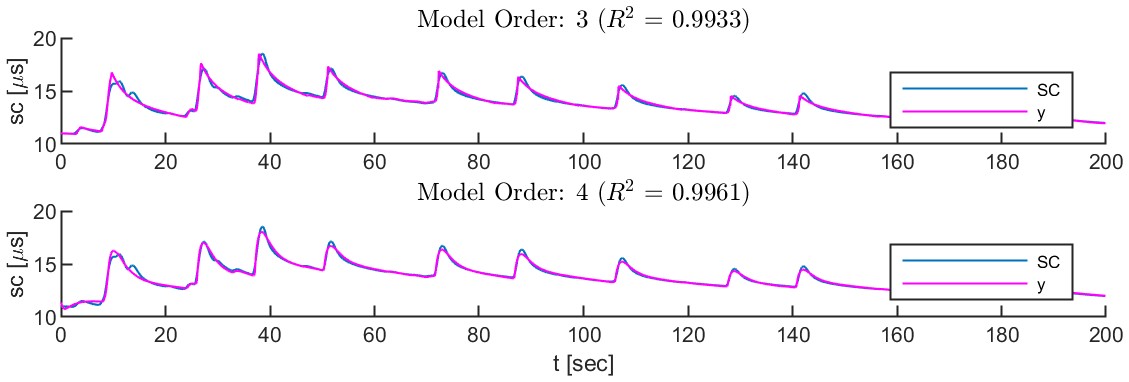}
\caption{Model Order Comparison}
\label{fig:PP88_ModelOrder}
\end{figure}

One could argue that lower-order models, like the third-order ones, also have sufficiently high average $R^2$. However, visual inspections reveal that the output of third-order models tend to have sharp peaks instead of the typical smooth peaks of SCRs, which can be viewed as underfitting. An example is shown in Figure \ref{fig:PP88_ModelOrder}. It can be seen that the fitting of the phasic response improves evidently from a third-order model to a fourth-order model, although the increase of $R^2$ is small. Based on all the results, we determined that the best model order for the proposed model is four.

\section{Decomposition of Skin Conductance}
\label{sec_model_decomp}

Decomposing sc into the tonic level and phasic response is a necessary step for obtaining key metrics of skin conductance. Improper decomposition may lead to inaccurate estimation of these metrics. Determining the tonic level and phasic response of skin conductance is challenging for several reasons. In psychophysiology, tonic level is defined as the background level in the absence of any internal or external stimuli \cite{stern2001}. According to this definition, tonic level can only be determined when no SCR is in progress. However, in most cases, it is not possible to determine the exact point in time when an SCR ends \cite{boucsein2012}. Traditionally, the tonic level is estimated by averaging SCR-free recording segments within a sufficient interval. The tonic level obtained by this approach is likely to be overestimated especially when the occurrence of SCR is frequent.

\subsection{Tonic Phasic Decomposition of Existing EDA Models}

Existing EDA models' assumptions and treatments of the tonic level are summarized in Table \ref{model_assumptions}. These treatments can be broadly categorized into three approaches.

\begin{figure}[!htbp]
\centering
\includegraphics[scale=0.2]{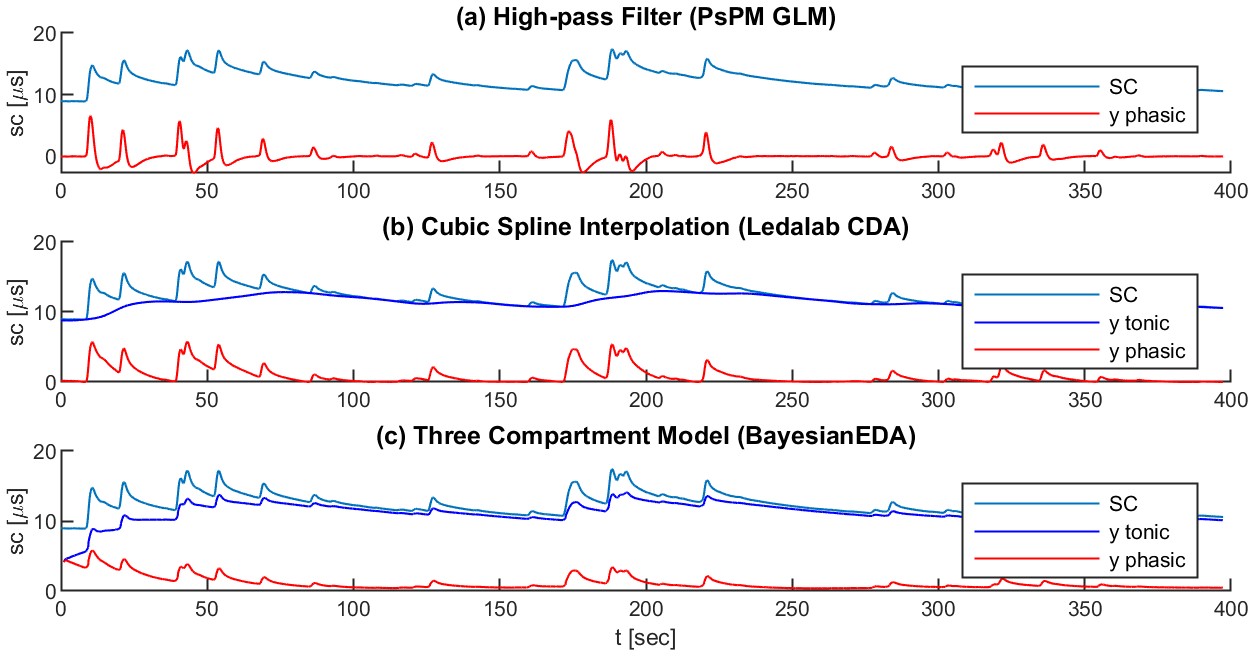}
\caption{Tonic-phasic Decomposition Approaches from Existing EDA Models}
\label{fig:TonicPhasic_methods}
\end{figure}

One approach involves removing the tonic level through the use of a high-pass filter, a method adopted by PsPM GLM and PsPM DCM. The distortion of the remaining phasic response is evident, as shown in subplot (a) of Figure \ref{fig:TonicPhasic_methods}. Finding a specific cutoff frequency to separate the tonic and phasic components is challenging because the skin conductance signal lacks inherent cycles. Instead, it features irregular events that correspond to internal and external stimuli. Moreover, the frequency bands of the tonic and phasic components likely overlap, given that the recovery of skin conductance responses typically merges smoothly with the underlying tonic level.

Another approach is to approximate the tonic level through the use of cubic splines.This is the most commonly used approach, adopted by several EDA models such as Ledalab CDA, Ledalab DDA, and cvxEDA. This method necessitates selecting a parameter: the spacing of the cubic splines' fixed points, which remains consistent throughout the signal's duration. Determining the value of this parameter can be challenging, as the appropriate value depends on the frequency at which SCRs occur, which is likely to vary. Moreover, due to the flexibility of cubic splines, the estimated tonic level often exhibits arbitrary fluctuations, as depicted in subplot (b) of Figure \ref{fig:TonicPhasic_methods}. These fluctuations may impact the accuracy of estimated skin conductance parameters, such as SCR amplitude.

A third approach, recently introduced in BayesianEDA, employs a three-compartment pharmacokinetic model. In this model, sweat in the duct (the first compartment) disperses into two compartments: the surface of the skin around the pore, producing the phasic response, and the coreneum, producing the tonic level. A simplified version of this model suggested equal division between the tonic level and phasic response, as depicted in subplot (c) of Figure \ref{fig:TonicPhasic_methods}. It appears that both the estimated tonic level and phasic response exhibit similar temporal dynamics in response to the stimuli. This result seems at odds with the definition of tonic level, which is expected to reflect the baseline skin conductance in the absence of stimuli.

\subsection{Skin Conductance Decomposition of UDM}

\begin{figure}[ht]
\centering
\includegraphics[scale=0.2]{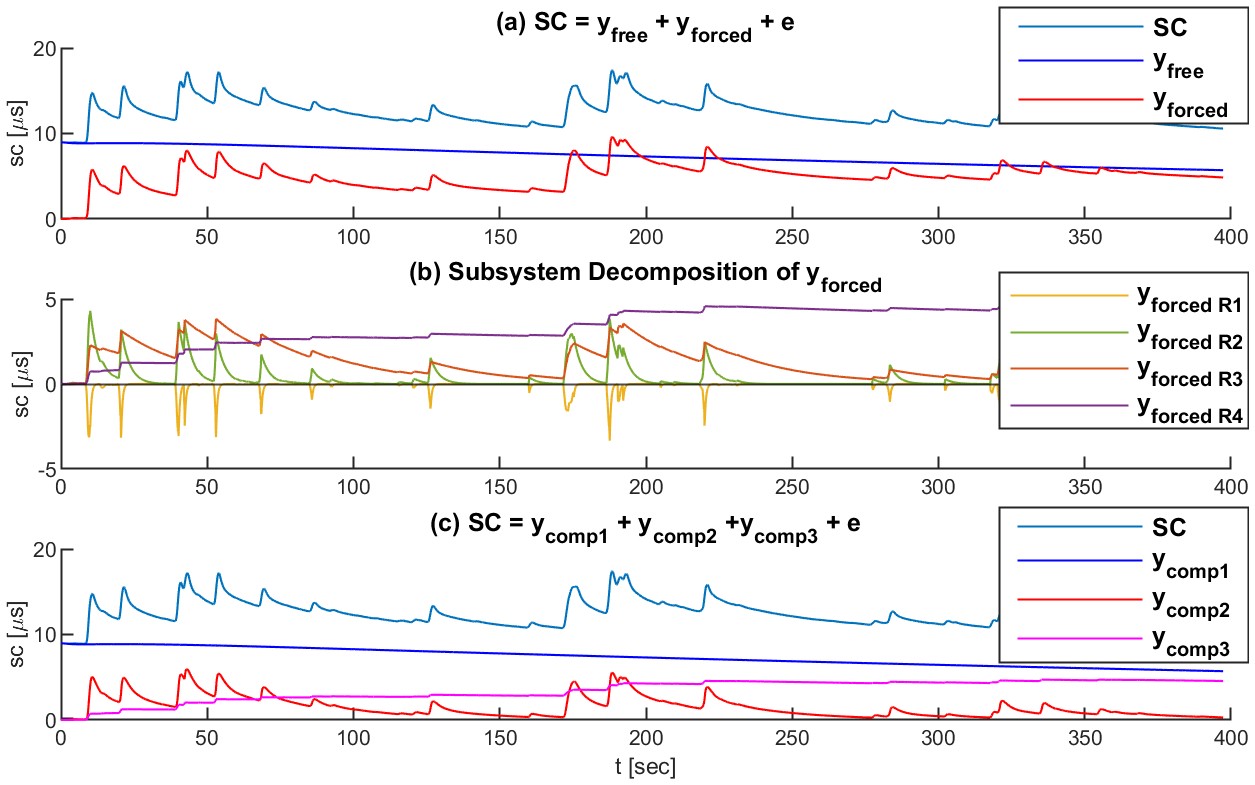}
\caption{UDM Skin Conductance Decomposition}
\label{fig:UDM_decomp}
\end{figure}

The proposed model UDM provides a unique perspective to the problem of tonic phasic decomposition. One of our model assumptions is that tonic level corresponds to the free response, while the phasic response corresponds to the forced response. Considering that the free response originates from the initial state, and the forced response is due to the input, this assumption in line with the definitions of the tonic level and the phasic response. It enables us to differentiate between the input and all other influencing factors. The estimated free and forced responses of a representative model are depicted in subplot (a) of Figure \ref{fig:UDM_decomp}. Notably, the free response exhibits a gradual decline. This is consistent with the observation that skin conductance tends to gradually decreases in the absence of stimuli, a pattern frequently seen in resting conditions.

The forced response can be further decomposed into the responses of subsystems. Using the Laplace transform and partial fraction decomposition, we can transform a fourth-order LTI system into the sum of four first-order subsystems. Consequently, the overall forced response is the sum of the responses from these subsystems, as depicted in subplot (b) of Figure \ref{fig:UDM_decomp}. Each first order subsystem corresponds to an exponential function in the time domain, as shown in Equation \ref{eq:exp_func}, where $g$ is the gain and $\tau$ is the time constant. The value of $\tau$ indicates the rate of change of the exponential function. The larger the time constant, the slower the rate of change. The four subsystems are referred to as $R_1$, $R_2$, $R_3$ and $R_4$, ranked in ascending order of their time constants. Subsystem $R_1$ has the smallest time constant and a negative gain, and therefore contributes to the fast rise of the SCRs. On the other hand, subsystems $R_2$, $R_3$ and $R_4$ all have positive gains, which contribute to the decline of the SCRs. It can be seen that the decline rate of $R_2$ is faster, while that of $R_3$ is slower. Together, they contribute to the gradual recovery of the SCRs. Subsystem $R_4$ has the largest time constant and its corresponding decline rate is so slow that it can be viewed as constant before the appearance of the next input. Consequently, the overall effect of $R_4$ manifests as a rising trend.

\begin{equation}
h(t) = g \cdot e^{-\frac{t}{\tau}}
\label{eq:exp_func}
\end{equation}

Based on the aforementioned analysis, we propose to decompose skin conductance into three components. The first component is the free response, because it is generated by initial state, which captures the contribution of past input and potentially other factors until the starting time. The second component is the forced response of the first three subsystems $R_1$, $R_2$ and $R_3$. This component aims to capture the short term electrodermal response to stimuli. The third component is the forced response of R4, which represents long term electrodermal response to stimuli. 

The proposed decomposition provides insights about the dynamics of skin conductance. Although the tonic level and phasic response are believed to have distinct physiological mechanisms and their definitions suggest independence, a correlation between empirically derived tonic metrics and phasic metrics is known to exist. The SCL obtained through the traditional method correlates strongly with the frequency of nonspecific SCR \cite{boucsein2012}. In addition, it's frequently observed that a series of SCRs within a short span can raise the SCL, a phenomenon often noted at the start of an experiment. If we treat the third component as part of the empirically determined tonic level, the correlation and the observation become understandable, as both the second and third components are generated by the input.

In summary, UDM decomposes skin conductance into three components, each with a unique role that offers a distinct contribution to the temporal dynamics. Importantly, this work does not attempt to propose new definitions for the tonic level and phasic response. Instead, our focus is on highlighting the discrepancy between theoretical and empirical definitions. Through our proposed three-component decomposition, we aim to offer a clearer understanding of this discrepancy and advocate for a more rigorous examination of skin conductance decomposition.

\section{Evaluation of the Estimated Sudomotor Nerve Activities}

One of the main applications of EDA models is to obtain an estimation of the sudomotor nerve activities (SNA). SNA is an important variable because it is indicative of physiological and psychological arousal, which makes it valuable for inferring affective states. In addition, obaining SCR parameters such as the amplitude using model based approaches requires the estimated SNA. However, the evaluation of the estimated SNA is challenging due to the difficulty of obtaining direct measurement of the ground truth. Although it is possible to measure skin sympathetic nerve activities directly through microneurography, the measurement typically contains aggregated activities of many nerve fibers including vasoconstrictor and sudomotor nerve fibers \cite{young2009}. Given a multi-unit microneurography recording, the composition and relative contribution of each fiber type is not identifiable \cite{greaney2017}. In addition, various biophysical processes happen around the skin and sweat glands that contribute to the change of electrical properties such as the skin conductance. The mechanisms of these processes are still far from being well understood.

Because of the reasons mentioned above, the evaluation of SNA remains an open question for discussion. Efforts to evaluate SNA have been made, providing valuable insights. Among these efforts, two approaches stand out as particularly inspiring. Bach et al. \cite{bach2014} proposed to evaluate the estimated SNA based on its sensitivity to differentiate experimental conditions that are known to be reliably associated with different levels of psychological arousal (e.g. aversive vs. neutral images). Hernando-Gallego et al. \cite{hernando2017} and Amin et al. \cite{amin2020} compared the sparsity levels of the estimated SNA from various models. In the following subsections, the estimated SNA from different EDA models will be compared based on these two perspectives. The EDA models considered in evaluation are the ones with executable codes released, which are Ledalab CDA \cite{benedek2010_CDA}, Ledalab DDA \cite{benedek2010_DDA}, PsPM GLM \cite{bach2010modelling}, PsPM DCM \cite{bach2010dynamic}, cvxEDA \cite{greco2015}, sparsEDA \cite{hernando2017}, and BayesianEDA \cite{amin2022}. The model proposed in this work is referred to as the unified dynamic model (UDM).  

A public dataset PsPM-trSP1 \cite{bach_data} was used for this evaluation. It contains skin conductance recordings of 60 participants (30 male, 30 female) viewing 45 neutral and 45 aversive images from the International Affective Picture Set (IAPS) \cite{lang1997}. The selected aversive and neutral images are based on normative ratings in IAPS. More details of the study can be found in \cite{bach2013improved}. Considering the length of the task and its repetitive nature, in order to mitigate the effects of habituation, only the first block of the task was used, which contains 15 aversive and 15 neutral images. Five participants whose skin conductance recordings have pronounced artifacts are excluded from the analysis, including 2 participants (ID: 17, 39) whose skin conductance have sudden large spikes that are likely to be movements artifacts, 2 participants (ID: 10, 18) whose recordings have high low signal-to-noise ratio and 1 participant whose skin conductance is negative (ID: 13). The remaining 55 participants were used in the analysis.

\subsection{Differentiating High and Low Arousal}

In this section, the sensitivity of the estimated SNA in differentiating high and low psychological arousal is compared, following the evaluation approach proposed by Bach et al. \cite{bach2013model, bach2014} This evaluation assumes physiological arousal accurately reflects psychological arousal. A general one-to-one relationship between psychological variables and physiological variables rarely exists. However, in the context of affective images, a relatively consistent relationship was found between the rated subjective psychological arousal of pictures in certain themes (e.g. threat and erotica) and the amplitudes of the ensuing SCRs \cite{bradley2001,bernat2006}.

The two conditions, aversive and netural images, correspond to high and low psychological arousal. In this context, psychological arousal is reflected by physiological arousal, which can be obtained through the estimated SNA. The hypothesis is that the mean arousal of viewing aversive images is larger than the mean arousal of viewing neutral images.

The Wilcoxon signed-rank test, a non-parametric version of the paired t-test, is used to test the hypothesis, because the distributions of the paired differences obtained from most of the models were positively skewed. The same preprocessing procedure described in Section \ref{sec_prep} was applied to the skin conductance recordings. 

For each EDA model mentioned in the beginning of this section, the average across-trial physiological arousal in response to aversive images and neutral images were computed for each participant. All the trials were used in the computation, regardless of whether an SCR occurred or not. Physiological arousal in this context is defined operationally as the sum of the estimated SNA within a latency window of 1-4 seconds after the display of the image \cite{benedek2010_CDA}. Because PsPM\_GLM does not have an estimated SNA, the estimated SCR amplitudes were used instead, following the procedure described in \cite{bach2014}.

\begin{table}[!ht]
\centering
\caption{Estimated Arousal in Response to IAPS Images}
\begin{tabular}{|c|c|c|c|c|}
\hline
Model & Aversive & Neutral & Test stat & p-value \\
\hline
Trough to Peak & 0.457 & 0.259 & 1198 & 0.000 \\
\hline
Ledalab CDA & 0.601 & 0.348 & 1369 & 0.000 \\
\hline
Ledalab DDA & 1.063 & 0.714 & 1301 & 0.000 \\
\hline
PsPM GLM & 1.635 & 0.571 & 1430 & 0.000 \\
\hline
PsPM DCM & 0.681 & 0.559 & 1284 & 0.000 \\
\hline
cvxEDA & 1.637 & 0.953 & 1254 & 0.000 \\
\hline
sparsEDA & 0.518 & 0.188 & 1181 & 0.000 \\
\hline
BayesianEDA & 8.293 & 4.981 & 1306 & 0.000 \\
\hline
UDM & 0.443 & 0.246 & 1333 & 0.000 \\
\hline
\end{tabular}
\label{table:SA}
\end{table}

The results of the Wilcoxon signed-rank test are shown in Table \ref{table:SA}. Although the ranges of the estimated arousals differ among the models due to the different procedures used, the ranges do not affect the result of the hypothesis testing. The Wilcoxon signed-rank tests have significant p-values (alpha=0.01) for all the models. It indicates that all the EDA models are able to differentiate high and low arousal under certain well-defined experimental conditions. Among them, PsPM\_GLM has the highest test statistic, and Ledalab\_CDA has the next highest. However, neither PsPM\_GLM nor Ledalab\_CDA enforce nonnegativity constraint on SNA. UDM has the highest test statistics among the models that ensure the estimated SNA are nonnegative. All the models except sparsEDA have higher test statistics than the traditional trough-to-peak approach.

\subsection{Evaluating the Sparsity Level of the Estimated Sudomotor Nerve Activities}
\label{sec_sna_sparsity}

\begin{figure}[!ht]
\centering
\includegraphics[scale=0.2]{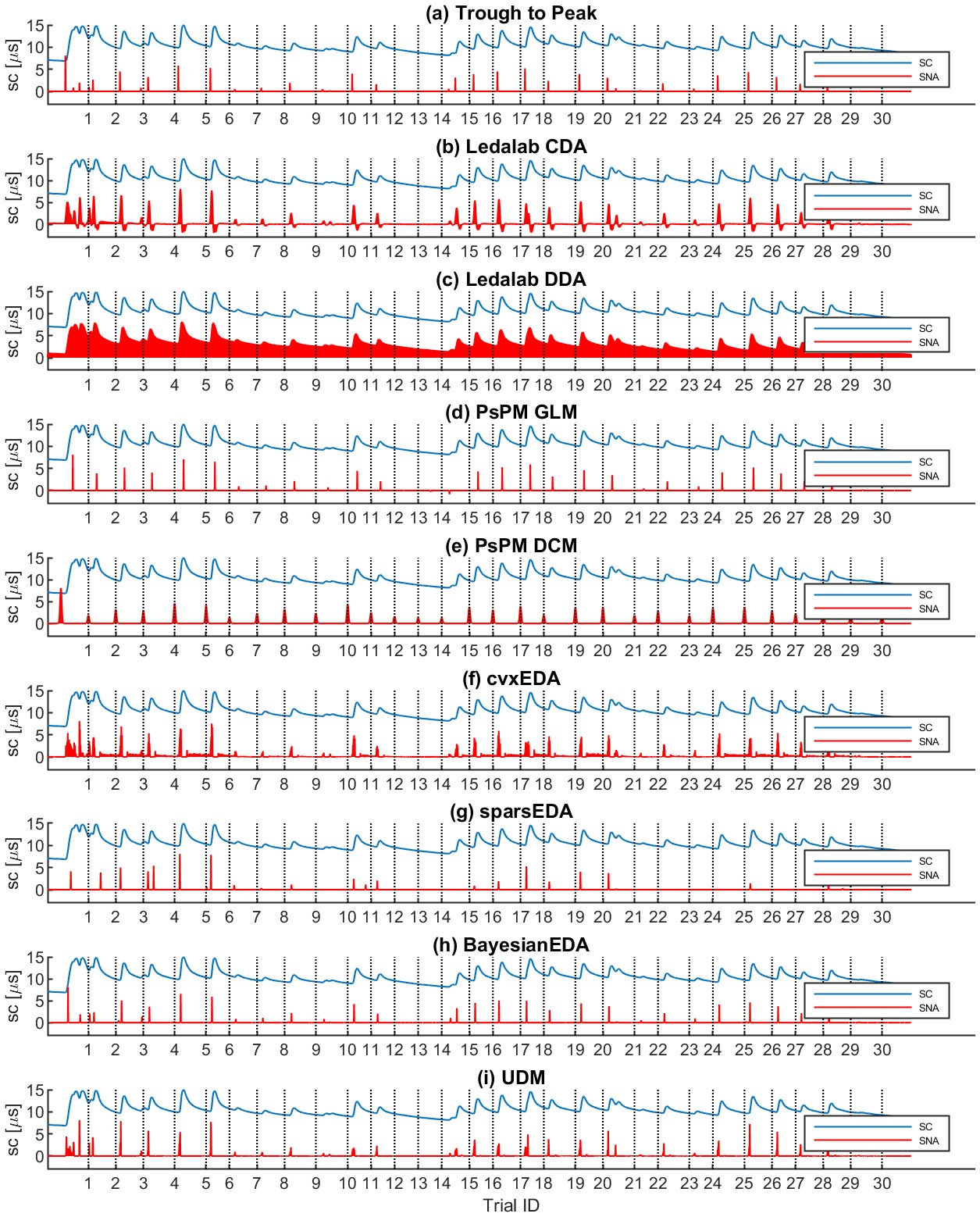}
\caption{Compare Estimated SNA from Different Models}
\label{fig:SNA_moderate}
\end{figure}

One of the assumptions made by almost all of the existing EDA models is that SNA signal is sparse. An important question is what the sparsity level of the estimated SNA should be, given a particular skin conductance recording. Lacking direct measurement of the ground truth, it is difficult to assess the appropriate sparsity level. However, it can be argued that there should be a one-to-one correspondence between each SNA burst and each SCR. Since SNA bursts are viewed as the direct cause of SCRs, it is expected that for each SCR there should be a corresponding SNA burst, and there shouldn’t be any SNA burst when there isn’t any SCR. Figure \ref{fig:SNA_moderate} shows a representative participant who has a moderate number of SCRs in response to the IAPS images. The SNAs were rescaled to facilitate the comparison of different models. The maximum SNA value after rescaling is equal to the range of the skin conductance recording, which is the difference between the minimum value and the maximum value. 

The estimated SNA of Ledalab CDA (subplot (b)) has negative portions right after each bursts. Negative values of SNA are physiologically implausible and violate the non-negativity assumption. It is likely caused by standard deconvolution with the bi-exponential function as the impulse response. Ledalab DDA (subplot (c)) has the densest SNA (nonzero throughout the whole segment) in this example, which clearly violates the sparsity assumption. 

PsPM GLM (subplot (d)) and PsPM DCM (subplot (e)) differ from other models in that SNA is estimated trial-by-trial. It can be seen that the SNA of PsPM GLM is negative for some of the trials (e.g. trial 14). This is because the tonic level was removed by a high pass filter in PsPM's preprocessing step, which distorts the shape of the remaining signal (presumably the phasic response). In PsPM DCM, the SNA for each trial was estimated using Bayesian inference. The SNA for a trial could be nonzero even when there is no SCR in response to it (e.g trial 12 and 13). This is because the posterior estimation is dominated by the prior estimation when there is no SCR observed.

The estimated SNA of cvxEDA (subplot (f)) is not sparse in that there is often a tail following each burst. The timings of SNA are expected to coincide with the onset of SCRs. However, the timing of these tails are around the recovery part of the SCRs, which is physiologically inexplicable. The possible reason might be the bi-exponential impulse response function is unable to fully account for the within-subject variation of the SCRs. On the other hand, the estimated SNA of sparsEDA (Figure \ref{fig:SNA_moderate} subplot (g)) is over-sparse. For example, there are no estimated SNA in correspondence to the SCRs of some trials (e.g. trial 24 and 26). 

Both BayesianEDA (subplot (h)) and UDA (subplot (i)) have the appropriate sparsity level in the sense that there is a one-to-one correspondence between each SNA burst and each SCR. In BayesianEDA, each individual impulse represents an SNA burst, while in UDM each impulse cluster represents an SNA burst. SNA bursts vary in duration as noted in \cite{bini1980, macefield1996}. In UDM, each estimated SNA burst consists of several adjacent impulses, which allows us to capture this variation.

In sum, among all the EDA models considered in this section, both UDM and BayesianEDA are able to consistently obtain the sparsity level of SNA bursts that matches the frequency of the SCRs. UDM's estimated SNA bursts vary in duration, which is more consistent with the observations of physiological study.

\section{Conclusions}

In this work, we proposed a fourth order LTI dynamic system for the modeling of the temporal dynamics of skin conductance, combining the dynamics of both the tonic level and the phasic response. The assumptions of the model are specified explicitly and rigorously. A variation of the EM-algorithm is used to estimate the model parameters and the latent variable, the sudomotor nerve activity signal. The model was applied to a large dataset (IASL\_ARB) that includes more than 200 participants. The majority ($95.5\%$) of the models have $R^2$ above 0.99. 

Another contribution of this work is that it provides unique insights into the temporal dynamics of skin conductance. We proposed a three-component decomposition of skin conductance: the contribution from the initial state, the short-term response to the input, and the long-term response to the input. This decomposition elucidates the discrepancy between the defined tonic level and its empirical estimation, and it is also able to explain the observed correlation between the empirical tonic level and phasic response.

Finally, we evaluated the estimated SNA and compared it to the ones obtained using other existing EDA models. The sensitivity of the estimated SNA in differentiating high and low arousal was evaluated using the Wilcoxon signed-rank test. UDM had the third highest test statistic among all the eight models. The two models that had higher test statistics (PsPM GLM and Ledalab CDA) did not enforce the nonnegativity constraint on SNA. Additionally, we evaluated the sparsity level of SNA under the assumption that each SNA burst should correspond one-to-one with each SCR. UDM consistently achieved an appropriate sparsity level for SNA bursts, aligning with the frequency of the SCRs.

\section{Discussions}

\subsection{Limitations}

While this research provides valuable insights, there are some limitations that need to be acknowledged. First, the model estimation algorithm employed here, a variant of the Expectation Maximization (EM) algorithm, requires refinement. The algorithm starts with an initial SNA estimate derived from peak detection. Although this generally offers a satisfactory initial approximation, challenges arise in segments with severe SCR overlap, making it difficult to distinguish between troughs and peaks. Consequently, this occasionally results in undesirable final estimates. For instance, sometimes the estimated free response might exhibit segments that are negative or exceed the actual skin conductance measurements. To mitigate this, one could consider imposing constraints on the initial state during model estimation. Nonetheless, efficiently and accurately determining the initial condition for dynamic systems remains a challenge in system identification \cite{kolic2022}.

Furthermore, UDM has been predominantly validated on skin conductance recordings procured under discrete stimuli, such as loud noises and IAPS images. The focus on such stimuli arises from their characterization as distinct and well-defined brief events, which can be modeled by Dirac delta functions. This representation simplifies the assessment of linearity and time-invariance properties. However, in the cases of continuous stimuli such as mental arithmetic or public speaking tasks, the current representation of the input and latent variables may not be appropriate. Addressing the limitations of the current model estimation algorithm is crucial before extending its application to more complex stimuli.

\subsection{Future Work}

An important next step of future work involves addressing the limitations of the current model estimation algorithm. One approach could be to initialize the EM algorithm with model parameters (such as time constants and gains) based on the ranges and distributions obtained in this work. This approach allows us to use multiple random initializations to select the best estimation and avoid getting stuck in undesirable local optima. Furthermore, incorporating constraints on the initial condition during system identification could be valuable to ensure that the decomposed components consistently stay between zero and the skin conductance measurement.

Another intriguing avenue to pursue is to delve deeper into individual differences. Physiological responses are known to exhibit considerable variance among individuals. This variability is clearly reflected in the range of estimated model parameters. However, similarities among individuals also exist. To explore and characterize these differences and similarities, we attempted several clustering techniques in this work, utilizing features extracted from the estimated model and the SNA. However no informative group patterns have yet been discovered. A potential reason could be the strong correlation amongst many features. Broadening the scope to include features from other physiological systems, such as the cardiovascular system, might enhance the clustering results.

\section*{Acknowledgment}

The authors would like to express their gratitude to Dr. Karen Quigley for providing the IASL\_ARB dataset. Containing a large number of participants, this dataset displays a rich diversity of individual differences, making it crucial for the development of a robust model in this research.

\bibliographystyle{unsrtnat}
\bibliography{references}  

\end{document}